\documentclass[prd,showpacs,preprintnumbers,amsmath,amssymb,superscriptaddress,floatfix,nofootinbib]{revtex4}
\usepackage{graphicx}
\usepackage{epstopdf}
\usepackage{amsmath}
\usepackage{amsfonts}
\usepackage{amssymb}
\usepackage{bm}

\begin{document}

\title{The $Z_{cs}(3985)$ as a threshold effect from the $\bar D_s^* D +\bar D_s D^*$ interaction}

\author{Natsumi Ikeno}
\email{ikeno@tottori-u.ac.jp}
\affiliation{Department of Agricultural, Life and Environmental Sciences, Tottori University, Tottori 680-8551, Japan}
\affiliation{Departamento de F\'{\i}sica Te\'orica and IFIC,
Centro Mixto Universidad de Valencia-CSIC Institutos de Investigaci\'on de Paterna, Aptdo.22085, 46071 Valencia, Spain}

\author{Raquel Molina}
\email{raquel.molina@ific.uv.es}
\affiliation{Departamento de F\'{\i}sica Te\'orica and IFIC,
Centro Mixto Universidad de Valencia-CSIC Institutos de Investigaci\'on de Paterna, Aptdo.22085, 46071 Valencia, Spain}

\author{Eulogio Oset}
\email{oset@ific.uv.es}
\affiliation{Departamento de F\'{\i}sica Te\'orica and IFIC,
Centro Mixto Universidad de Valencia-CSIC Institutos de Investigaci\'on de Paterna, Aptdo.22085, 46071 Valencia, Spain}
\date{\today}

\begin{abstract}
We study the $e^+ e^- \to K^+ ( D_s^{*-} D^0 + D_s^- D^{*0})$ reaction recently measured at BESIII, from where a new $Z_{cs}$ state has been reported. We study the interaction of $\bar D_s D^*$ with the coupled channels $J/\psi K^-$, $K^{*-} \eta_c$, $D_s^- D^{*0}$, $D_s^{*-} D^0$ by means of an extension to the charm sector of the local hidden gauge approach. We find  that the  $D_s^- D^{*0}+ D_s^{*-} D^0$ combination couples to $J/\psi K^-$ and  $K^{*-} \eta_c$, but the $D_s^- D^{*0}- D_s^{*-} D^0$ combination does not. The coupled channels help to build up strength in the $D_s^- D^{*0}+ D_s^{*-} D^0$ diagonal scattering matrix close to threshold and, although the interaction is not strong enough to produce a bound state or resonance, it is sufficient to produce a large accumulation of strength at the $\bar D_s D^*$ threshold in the $e^+ e^- \to K^+ ( D_s^{*-} D^0 + D_s^- D^{*0})$ reaction in agreement with experiment.
\end{abstract}

\maketitle
\section{Introduction}

The discovery of the $Z_c(3900)$ at Ref.~\cite{beszc} and Ref.~\cite{bellezc} was a turning point in hadron spectroscopy since it provided a clear example of an exotic meson state. Many possible types of hadronic structures were proposed, from compact tetraquarks \cite{marina,agaev} to hadronic molecules \cite{guojuan} and even a threshold effect \cite{swanson}, although challenged in Refs.~\cite{hanhart,pilloni}. Subtle points of the interaction, as the small attractive $D\bar{D}^{*}$ interaction in one channel for $I=1$ \cite{aceti}, led gradually to an interpretation as some type of molecular state, slightly bound or virtual, helped by coupled channels~\cite{aceti,he,juanmiguel,nefediev,guowang,qwang,entem,ikeda,sasa,hechen,chenchen,meng}.

In Ref.~\cite{aceti} a barely bound state or a virtual state of $\bar D^* D +\bar D D^*$ nature were found, which produced peaks a few MeV above threshold in the $D\bar{D}^{*}$ mass distribution of the  $e^{+}e^{-} \to \pi D\bar{D}^{*}$ reaction measured at BESIII \cite{Ablikim:2013xfr}. Such a situation is quite common and was also stressed in Ref.~\cite{juanmiguel}.

 In a very recent experiment the BESIII collaboration has reported a state, named $Z_{cs}(3985)$, from a peak near threshold  in the $ D_s^{*-} D^0 $, $ D_s^- D^{*0}$ invariant mass distribution of the $e^+ e^- \to K^+ ( D_s^{*-} D^0 +  D_s^- D^{*0})$ reaction \cite{newbes}. The mass and width reported are
\begin{eqnarray}
M &=& 3982.5^{+1.8}_{-2.6} \pm 2.1  \ {\rm MeV}, \\
\Gamma &=& 12.8^{+5.3}_{-4.4} \pm 3.0  \  {\rm  MeV}.
\end{eqnarray}

The state lies about 7 MeV above the $ D_s^{*-} D^0 $, $ D_s^- D^{*0}$ threshold and could be the SU(3) partner of the $Z_c(3900)$ state, where a $u$ or $d$ quark has been replaced by an $s$ quark. 

   The reaction from the theoretical community to that finding has been fast. Several works have proposed different explanations for the new state.  Prior to the experiment, a QCD sum rule calculation with the $\bar D_s^* D +\bar D_s D^*$ molecular configuration produced a state with a mass of $3960$ MeV and  uncertainty about $100$ MeV~\cite{nielsen}. In Ref.~\cite{wanqiao}, QCD sum rules are also applied to the investigation of $\bar K^* \eta_c$, $\bar D_s^* D$, $\bar D_s D^*$ configurations and also a compact tetraquark state, with uncertainties of about 150~MeV in the mass. Another work based on sum rules has been done in Ref.~\cite{zhigang} establishing a possible tetraquak state with these quantum numbers and a mass compatible with experiment within the typical uncertainties in the mass of the sum rules, and a similar one is done in Ref.~\cite{Sungu:2020zvk} investigating also the resonance properties at finite temperature. 
Work along similar lines is also done in Ref.~\cite{azizi} making extrapolations to the beauty sector.
Quark model calculations have also been reported in Ref.~\cite{ping} considering two types of structures: meson-meson and diquark-antidiquark, favoring the latter option.
In Ref.~\cite{shilinmeng} arguments of SU(3) symmetry are used to relate the $Z_c(3900)$, assumed to be a molecular state of positive $G$-parity $\bar D^* D +\bar D D^*$, 
  to  a state made of $\bar D_s^* D +\bar D_s D^*$. The state obtained is the U-spin partner of $Z_c(3900)$ as a resonance within
coupled-channel using  SU(3)$_F$  and heavy quark spin symmetry (HQSS).
The coupled channels  considered are $J/\psi \bar K $,  $\frac{1}{\sqrt 2}(\bar D_s^* D +\bar D_s D^*)$ and $\bar D_s^* D^*$. In a follow up work of this idea, higher orders in the interaction are considered in Ref.~\cite{shilinwang} and predictions are made for the beauty sector.
A different proposal is made in Ref.~\cite{xiangliu}
where the authors suggest that the $Z_{cs}(3985)$ can be explained as
a reflection structure of the charmed-strange meson $D_{s2}^*(2573)$, which is produced from the open-charm decay of Y(4660) with a $D^*_s$ meson. In Ref.~\cite{juanguo}, with the input of experiment from the $Z_c(3900)$ and $Z_c(4020)$ masses, and using arguments of SU(3) and heavy quark spin symmetry, a state made of $\bar D_s^* D$, $\bar D_s D^*$ is obtained, which can be associated to the newly found $Z_{cs}$ state, while another one made of $D^* \bar D_s^*$ is predicted. The one boson exchange picture is used in Ref.~\cite{riuchen}, where considering only the 
$\bar D_s^* D$, $\bar D_s D^*$ and $D^* \bar D_s^*$ configurations no resonant state can be found. A different line of research is followed in Ref.~\cite{qiangzhao} where the authors  parametrize
the reaction mechanisms  by $S$ and $D$-wave production amplitudes in the  $e^+ e^- \to \pi D \bar D^*$ and $e^+ e^- \to  K \bar D_s D^*$ reactions, largely constrained by
the Jackson angular distributions, in terms of the recoiled pion or kaon in the final states. Then the lineshapes of the invariant mass spectra provide constraints on the pole structures of the $Z_c(3900)$ and $Z_{cs}(3982)$, which appear as either virtual or bound states. Finally, in Ref.~\cite{chuwen} the molecular picture is again exploited using the 
$D_s^{(*)-} D^{(*)0}$ components and Lagrangians which describe the
interaction of charmed mesons, taking the chiral and hidden local symmetries
into account.

  Our picture has the closest resemblance to Ref.~\cite{shilinmeng}. We consider the same coupled channels but we also include the $ K^{*-} \eta_c$ channel. We ignore the mixing with the vector vector channel, which can give rise to new states~\cite{shilinmeng,juanguo}, but mix poorly with the $\bar D_s^* D$, $\bar D_s D^*$ channels close to their threshold, since they involve anomalous couplings with small three momenta. As we shall see, we do not need to assume the $\bar D_s^* D +\bar D_s D^*$ structure instead of the $\bar D_s^* D -\bar D_s D^*$ one, as done in Ref.~\cite{shilinmeng}, since it comes directly from the theory. Here we do a quantitative evaluation of the transition potentials between the channels, using an extension of the local hidden gauge approach~\cite{hidden1,hidden2,hidden4,nagahiro}, while these potentials are parametrized in Ref.~\cite{shilinmeng} in terms of four free parameters with constrains from HQSS. 
   
    We show that even not getting a bound state below threshold, the strength of the interaction creates a structure in the $e^+ e^- \to  K \bar D_s  D^*$ reaction above and very close to the $\bar D_s  D^*$ threshold compatible with the peak  observed in the experiment. 

\section{Formalism}\label{sec:form}
We take the vector-pseudoscalar channels ($VP$)
\begin{equation}
 J/\psi K^-\, (1), \ \  K^{*-} \eta_c \, (2), \ \ D^{*-}_s D^0\, (3), \ \ D^{-}_s D^{*0}\, (4),
\label{eq:channel}
\end{equation}
and following \cite{aceti} we study the interaction between these channels using the extension of the local hidden gauge approach~\cite{hidden1,hidden2,hidden4,nagahiro} that was used in Ref.~\cite{aceti}. The peculiar feature of this interaction, based on the exchange of vector mesons, is that one cannot exchange light vectors with the quark configuration of those states, which renders the strength of the interaction very small in the present case.  Hence, as in the case of the $Z_c(3900)$, we expect to obtain only very weakly bound states or virtual states. In a general case one can exchange light vectors and the interaction is strong enough to generate axial vector mesons. This is the case for the light axial vector mesons, $a_1(1260)$, $b_1(1235)$, $h_1(1170)$, $h_1(1380)$, $f_1(1285)$ and two $K_1(1270)$, which are generated from the interaction of light $VP$ coupled channels~\cite{lutz,luis,rocageng}. In Refs.~\cite{luis,rocageng} chiral Lagrangians for the $VP$ interaction were used from Ref.~\cite{birse}, which are known to be equivalent to the results from the exchange of vector mesons with the local hidden gauge Lagrangians~\cite{rafael} (practical explicit examples of the equivalence can be seen in Ref.~\cite{debastiani}). 

\begin{figure}[tb!]
  \centering
  \includegraphics[width=0.45\textwidth]{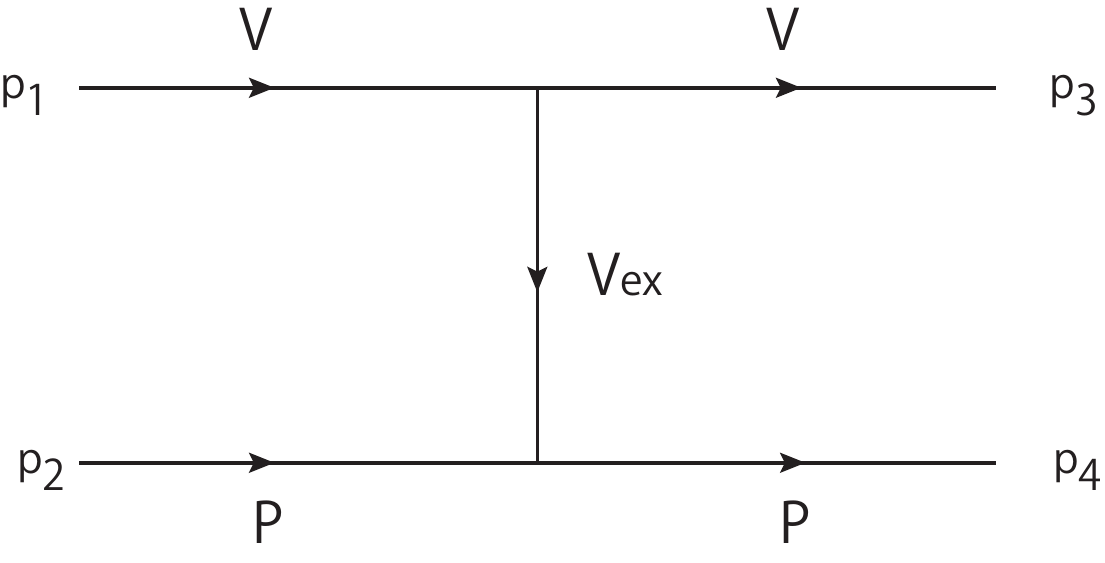}
  \caption{ Diagrammatic mechanism for the $VP \to VP$ interaction through vector meson exchange. }
  \label{fig:1}
\end{figure}

   The interaction proceeds as shown in Fig.~\ref{fig:1}. 

      We need the local hidden gauge Lagrangians $L_{VVV}$ and $L_{VPP}$, which, as in Ref.~\cite{aceti}, are extended to the charm sector using
\begin{equation}
 \mathcal{L}_{VPP}=-ig\langle V^{\mu}[P,\partial_{\mu}P]\rangle\ ,
\label{eq:VVPlag}
\end{equation}
\begin{equation}
 \mathcal{L}_{VVV}=ig\langle (V^{\mu}\partial_{\nu}V_{\mu}-\partial_{\nu}V_{\mu}V^{\mu})V^{\nu}\rangle\ ,
\label{eq:VVVlag}
\end{equation}
with $g=M_V/2f$ ($M_V=$800~MeV, $f=$93~MeV), and 
\begin{equation}
P=\left(
\begin{array}{cccc}
\frac{\eta}{\sqrt{3}}+\frac{\eta'}{\sqrt{6}}+\frac{\pi^0}{\sqrt{2}} & \pi^+ & K^+&\bar{D}^0\\
\pi^- &\frac{\eta}{\sqrt{3}}+\frac{\eta'}{\sqrt{6}}-\frac{\pi^0}{\sqrt{2}} & K^{0}&D^-\\
K^{-} & \bar{K}^{0} &-\frac{\eta}{\sqrt{3}}+\sqrt{\frac{2}{3}}\eta'&D^-_s\\
D^0&D^+&D^+_s&\eta_c
\end{array}
\right)\ ,
\label{eq:pfields}
\end{equation}

\begin{equation}
V_\mu=\left(
\begin{array}{cccc}
\frac{\omega}{\sqrt{2}}+\frac{\rho^0}{\sqrt{2}} & \rho^+ & K^{*+}&\bar{D}^{*0}\\
\rho^- &\frac{\omega}{\sqrt{2}}-\frac{\rho^0}{\sqrt{2}} & K^{*0}&D^{*-}\\
K^{*-} & \bar{K}^{*0} &\phi&D^{*-}_s\\
D^{*0}&D^{*+}&D^{*+}_s&J/\psi
\end{array}
\right)_\mu\ .
\label{eq:vfields}
\end{equation}

One observation must be done concerning Eq.~(\ref{eq:VVVlag}). We are interested in the situation close to threshold involving $D_s^*$ and $D^{*0}$ vectors, which will then have very small three momenta. In this case we ignore the $\epsilon^0$ component of the polarization vector for the external vectors. This means that $V^{\nu}$ of Eq.~(\ref{eq:VVVlag}) cannot be an external vector because then $\nu=1,2,3$ and $\partial_{\nu}$ will give three momenta $\vec p$ which are neglected in this case. Hence,  $V^{\nu}$ can only be the exchanged vector in Fig.~\ref{fig:1} and then the analogy between Eqs.~(\ref{eq:VVPlag}) and (\ref{eq:VVVlag}) is manifest. The apparent different sign also disappears since  $\epsilon^{\mu}\epsilon'_{\mu}= - \vec \epsilon \cdot \vec \epsilon~'$ for the external vectors, and thus, the vertices are formally identical replacing the external pseudoscalars by vector mesons in Eqs.~(\ref{eq:pfields}), (\ref{eq:vfields}). 

   The interaction between the channels $i, j$ is given by 
\begin{equation}
 V_{ij} = C_{ij} g^2 (p_2 + p_4) (p_1 + p_3),
\end{equation}
which upon projection in $s$-wave reads as
\begin{equation}
  V_{ij} =  g^2 C_{ij} \frac{1}{2} \left[ 3 M^2_{12} - (m^2_1 +m^2_2 + m^2_3 + m^2_4 ) - \frac{1}{M^2_{12}} (m^2_1 - m^2_2) (m^2_3 - m^2_4) \right] , 
\end{equation}
where $M_{12}^2=(p_1+p_2)^2$ and the masses correspond to the states in the order depicted in Fig.~\ref{fig:1}. The matrix $C_{ij}$ is shown below ($C_{ji}=C_{ij}$)
\begin{equation}
\label{eq:cmatrix}
C_{ij}=\left( \begin{array}{cccc}
0 \ \ \ & \ 0   & \frac{1}{m^2_{D^*}} \ \  &  \frac{1}{m^2_{D^*_s}} \\
  \ \ &  \ 0 \  & \frac{1}{m^2_{D^*_s}} \  \  & \frac{1}{m^2_{D^*}} \\
  \ \ &  \  \  & - \frac{1}{m^2_{J/\psi}}   \  \  & 0 \\
 \ \ \ &      & \ \ \  & - \frac{1}{m^2_{J/\psi}} \end{array} \right)\ ,
\end{equation}
where the product $\vec \epsilon \cdot \vec \epsilon~'$ has been omitted since it factorizes in the Bethe-Salpeter equation and plays no role in the discussion.

   In Eq.~(\ref{eq:cmatrix}) we have neglected $q^2$ in $(q^2-M_{D^*}^2)^{-1}$ of the vector propagator. Later on we shall implement small corrections to account for $q^2$. To be able to exploit the symmetry of this interaction we shall use average $D$ and $D_s$ and 
$D^*$ and $D_s^*$ masses as 
\begin{equation}
  \bar m_{D} = 1916~ {\rm MeV}, \ \ \bar m_{D^*} = 2060~ {\rm MeV}.
\end{equation}

In this case it is convenient to work with the combination of states
\begin{eqnarray}
 A &=& \frac{1}{\sqrt{2}} ( D^-_s D^{*0} + D^{*-}_s D^0 ), \\
 B &=& \frac{1}{\sqrt{2}} ( D^-_s D^{*0} - D^{*-}_s D^0 ).
\end{eqnarray}

We can immediately see that the combination $A$ couples to $J/\psi K^-$ and $K^{*-} \eta_c$, while the combination $B$ does not couple. Then the $B$ combination stands as a single channel with a  very weak interaction mediated by $J/\psi$ exchange which does not lead to any bound states. The $A$ component gets reinforced by a factor $\sqrt 2$ in its transition to  $J/\psi K^-$ and $ K^{*-} \eta_c$ and this nondiagonal transition is much stronger than the diagonal one, making the presence of the coupled channels most welcome to eventually produce a weakly bound state. 

  The analogy to the $Z_c$ case is clear. In Ref.~\cite{aceti} it was also found that the $\bar D D^*- \bar D^* D$ combination did not bind, while the $\bar D D^*+ \bar D^* D$ combination produced weakly bound states or virtual ones. The $\bar D D^*+ \bar D^* D$ combination in $I=1$ has $G$-parity
 positive and its association to the $Z_c$ is natural. In the present case there is no $G$-parity but still the $\bar D_s D^*+ \bar D^*_s D$ combination is the only one that can produce a weakly bound state. In fact, this is the mixture that was assumed in Ref.~\cite{shilinmeng} by analogy to the positive $G$-parity state in the $Z_c$ case. Within our approach this combination appears as a consequence of the interaction of the extended local hidden gauge approach. In view of this, we take now the states 
\begin{equation}
 J/\psi K^- ~(1), \ \  K^{*-} \eta_c~(2), \ \  \frac{1}{\sqrt{2}} ( D^-_s D^{*0} + D^{*-}_s D^0 )~(3)
\end{equation}

and the new matrix $C_{ij}$ reads as shown below ($C_{ji}=C_{ij}$) 
\begin{equation}
\label{eq:cmatrix2}
{C}_{ij}=\left( \begin{array}{ccc}
0 \ \ \ & \ 0 \  & \frac{ \sqrt{2}}{\bar m^2_{D^*}}  \\
  \ \ &  \ 0 \  & \frac{ \sqrt{2} }{\bar  m^2_{D^*}} \\
  \ \ &  \  \  & - \frac{1 }{m^2_{J/\psi}}   \end{array} \right)\ .
\end{equation}

Since we are interested in invariant masses of $\bar D_ s D^*$ close to threshold, it is easy to take into account the $q^2$ correction in the propagators $(q^2-M_{D^*}^2)^{-1}$. We find $q^2=0$ for the diagonal transition and for $V_{13}$, $V_{23}$ 
\begin{equation}
 q^2 = M^2 + M^2_{D^*} - 2 E  M^2_{D^*},
\end{equation}
with $M=m_{J/\psi}$ for $V_{13}$ and $M=m_{K^{*-}}$ for $V_{23}$  and 
\begin{equation}
 E = \frac{M^2_{12} + M^2 - m^2}{2M_{12}},
\end{equation}
with $m=m_K$ for $V_{13}$ and $m=m_{\eta_c}$ for $V_{23}$. This correction increases the denominators $1/m_{D^*}^2$ in Eq.~(\ref{eq:cmatrix2}) by about 20\% and we take it into account.

   We then solve the Bethe-Salpeter equation in coupled channels 
\begin{equation}
T =\left[1-VG\right]^{-1} V,
\label{eq:BS}
\end{equation}
where $G$ is the diagonal vector-pseudoscalar loop function which we evaluate with cut off regularization as done in Ref.~\cite{aceti}. Then, 
\begin{equation}
\label{eq:1striemann}
G_l=\int\frac{d^3q}{(2\pi)^3}\,\frac{\omega_1+\omega_2}{2\omega_1\omega_2}\,\frac{1}{(P^0)^2-(\omega_1+\omega_2)^2+i\epsilon}\ ,
\end{equation}
with $\omega_1=\sqrt{m^2+\vec{q}^{\ 2}}$, $\omega_2= \sqrt{M^2+\vec{q}^{\ 2}}$ ($m$, $M$ the pseudoscalar and vector masses of the $l$ channel), and $|\vec q| < q_{\rm max}$. The value of $ q_{\rm max}$ was taken around 700--850~MeV in Ref.~\cite{aceti}.  

\section{results}
\begin{figure}[tb!]
  \centering
  \includegraphics[width=0.7\textwidth]{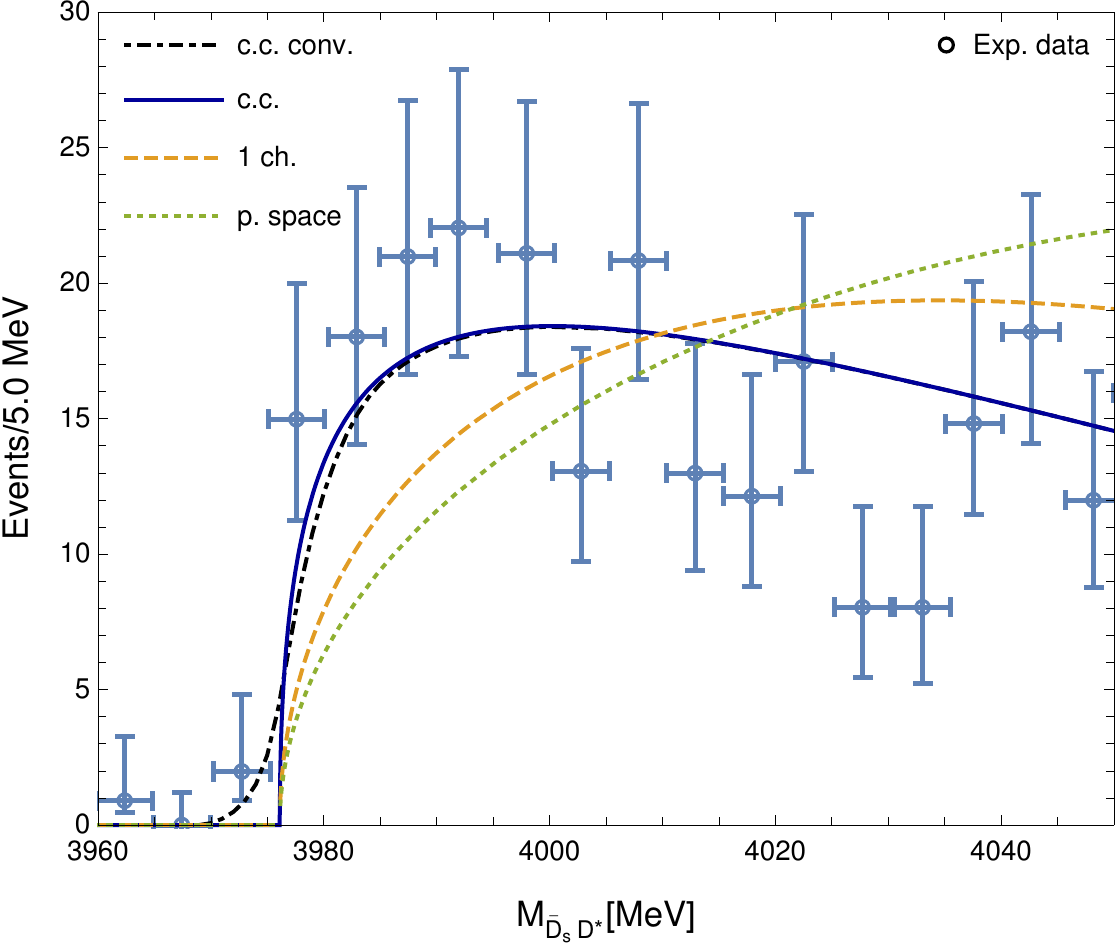}
  \caption{ Results of ${d\sigma}/{d M_{\bar{D}_s D^*}}$. Solid line: Result for the $\bar D_s D^{*} + \bar D^{*}_s D $ combination with its coupled channels (c.c.). Dashed line: Result for the single channel $\bar D_s D^{*} - \bar D^{*}_s D $ combination (1~ch.). Dotted line: phase space. Dashed-dotted line: result folded with the experimental resolution (c.c.~conv.). The data are taken from Ref.~\cite{newbes} for $\sqrt{s} = 4681$~MeV.
 }
  \label{fig:2}
\end{figure}

As in Ref.~\cite{aceti} we take the differential cross section as
\begin{equation}
\frac{d\sigma}{d M_{\bar{D}_s D^*}} =  \frac{1}{s\sqrt{s}} \ p \tilde{q} \ N  \ \left|T_{33}\right|^2 ,
\label{dsigma}
\end{equation}
with $\sqrt{s}$ the $e^+ e^-$ center of mass energy and
\begin{align}
 p&=\frac{\lambda^{1/2}(s,m^2_K, M^2_{\bar{D}_s D^*})}{2\sqrt{s}},\\
\tilde{q}&=\frac{\lambda^{1/2}(M^2_{\bar{D}_s D^*},m^2_{D_s},m^2_{D^*})}{2 M_{\bar{D}_s D^*}},
 \end{align}
and $N$ a normalization constant, neglecting a background since in the region of small invariant mass $M_{12} $ close to threshold our amplitude $T_{33}$ should account for all the strength of the interaction, 
and compare the results with an arbitrary normalization $N$ with the data of Ref.~\cite{newbes}.
Within the range of cut off 700--850~MeV used in Ref.~\cite{aceti} we find a fair agreement with the data. We, however, take advantage of the fact that to obtain Eq.~(\ref{eq:cmatrix2}), using the matrices of Eqs.~(\ref{eq:pfields}), (\ref{eq:vfields}), an explicit breaking of SU(4) symmetry has been implemented by the extended local hidden gauge approach when using physical mass $m_{D^*}$ and $m_{J/\psi}$ for the exchange of vector mesons.
A different pattern of SU(4) breaking was used in Ref.~\cite{aceti} following Ref.~\cite{gamermann} and the diagonal term in Eq.~(\ref{eq:cmatrix2}) had a factor $\displaystyle \psi=-\frac{1}{3}+\frac{4}{3}\left(\frac{m_L}{m_H}\right)^2$ instead of
$\displaystyle \left(\frac{m_V}{m_{J/\psi}}\right)^2$ with the $m_L$, $m_H$, the light and heavy masses used there, which increases the strength of this diagonal channel by a factor around two. To keep the closest analogy to Ref.~\cite{aceti}, we also take this factor and we present the results multiplying the diagonal term in Eq.~(\ref{eq:cmatrix2}) by $2.8$, which still gives a very weak interaction term, and use $q_{\rm max} = 750$~MeV in the middle of the range of Ref.~\cite{aceti}. The results are shown in Fig.~\ref{fig:2} (solid line) and we consider the agreement with the data sufficiently good. The convolution of the results with the experimental resolution is also done to get results for the lowest energies (dashed-dotted line).

Note that, unlike other works on this problem that look for bound states or resonances to compare with experiment, we do not follow that approach, but compare directly our results for the $\bar D_s D^{*}$ mass distribution with experiment. Even if the interaction is too weak to produce bound states or resonances, it can have some repercussion on threshold mass distributions. Thus we simply evaluate the $t$-matrix in our theory and see the observable effects in the threshold mass distribution.

We should also note that with this interaction we have not obtained a bound state, but a very strong cusp structure in the threshold of $\bar D_s  D^*$ that could qualify as a virtual state. However, to give significance to the results obtained we show in the same figure the results that we obtain with just phase space normalized to the same area below the curve (dotted line).
It is clear that the interaction used has the effect of accumulating strength close to threshold and produce agreement with the data. In the same figure we also show, normalized to the same data, the results obtained for the $\bar D_s D^{*} - \bar D^{*}_s D $ combination (dashed line), where only this channel with the diagonal interaction of Eq.~(\ref{eq:cmatrix2}) is operative. 
The outcome in this case does not differ much from that of the phase space and is clearly incompatible with the data. We should also mention that there can be dynamical reasons on why the $\bar D_s D^{*} - \bar D^{*}_s D $ combination is not produced, or only weakly in this reaction.
In the case of $\bar D D^{*} + cc$ with $I=1$ in Ref.~\cite{aceti}, this state has negative $C$-parity, which matches with the photon from $e^+ e^- $, and the $\bar D D^{*} - cc$ state would not be produced. In the present case, taking the $K^+ J/\psi K^-$ as a possible doorway for the reaction, it matches with the $C$-parity negative and can be produced, and then couple to the $\bar D_s D^{*} + \bar D^{*}_s D $ combination that we have studied, but not to $\bar D_s D^{*} - \bar D^{*}_s D $, as we have seen in Sec~\ref{sec:form}.

\section{Conclusions}
We have done a study for the $e^+ e^- \to K^+ ( D^{*-}_s D^0 + D^{-}_s D^{*0}) $  reaction taking into consideration the interaction of the coupled channels of Eq.~(\ref{eq:channel}) using for it an extension of the local hidden gauge approach to the charm sector. We saw that the $ J/\psi K^-$ and $K^{*-} \eta_c$  channels coupled only to the $(D^{*-}_s D^0 + D^{-}_s D^{*0})$ combination but not to the $(D^{*-}_s D^0 - D^{-}_s D^{*0})$ one. As a consequence of the coupled channels, the $( D^{*-}_s D^{0} + D^{-}_s D^{*0})$ combination developed much strength in the scattering matrix of 
$  D^{*-}_s D^{0} + D^{-}_s D^{*0} \to  D^{*-}_s D^{0} + D^{-}_s D^{*0}$,
leading to a strong cusp at the $\bar D_s D^*$ threshold, reminiscent of a failed bound or virtual state, which was, however, sufficient to produce a large accumulation of strength at the $\bar D_s D^*$ threshold in the $e^+ e^- \to K^+ ( D_s^{*-} D^0 + D_s^- D^{*0})$ reaction. This enhancement of the mass distribution around threshold was sufficient to produce agreement with experiment. We also showed the results that we  would obtain with the $D_s^- D^{*0}- D_s^{*-} D^0$ combination, which differed little from phase space and manifestly disagreed with the data. The exercise showed the virtue of the coupled channels in building up the strength of the $D_s^- D^{*0}+ D_s^{-*} D^0$ diagonal scattering matrix close to threshold.

The picture that we have does not anticipate a clear peak in the $J/\psi K^-$ and $K^{*-} \eta_c$ distributions, but we expect a clear cusp effect at the $\bar D_s D^*$ threshold as a consequence of the opening of this channel.

  In summary, although our picture does not produce a bound state nor resonance, we show that the interaction of the coupled channels, even if relatively weak, has sufficient strength to produce the enhancement close to threshold seen in the $e^+ e^- \to K^+ ( D_s^{*-} D^0 +  D_s^- D^{*0})$ reaction.

\section*{ACKNOWLEDGEMENT}
The work of N. I. was partly
supported by JSPS Overseas Research Fellowships and JSPS KAKENHI Grant Number JP19K14709. R. M. acknowledges support from the CIDEGENT program with Ref. CIDEGENT/2019/015 and from the spanish
national grant PID2019-106080GB-C21. This work
is partly supported by the Spanish Ministerio de Economia y Competitividad and European FEDER funds under
Contracts No. FIS2017-84038-C2-1-P B and No. FIS2017-84038-C2-2-P B. This project has received funding from
the European Unions Horizon 2020 research and innovation programe under grant agreement No 824093 for the
**STRONG-2020 project.

\end{document}